\documentclass[aps,showpacs,PRB,twocolumn,superscriptaddress,floatfix]{revtex4-2}

\usepackage[english]{babel}
\usepackage{amsmath}
\usepackage{graphicx} % Required for inserting images

\begin{document}

\title{Winding-Sector Transitions and Thermodynamic Incommensurability in Helical Valence Bond Phase under Tilted Boundary Conditions}

\author{Yan Liu}
\affiliation{Department of Physics and State Key Laboratory of Surface Physics, Fudan University, Shanghai 200438, China}

\author{Jie Lou}
\email{louejie@fudan.edu.cn}
\affiliation{Department of Physics and State Key Laboratory of Surface Physics, Fudan University, Shanghai 200438, China}

\author{Yan Chen}
\email{yanchen99@fudan.edu.cn}
\affiliation{Department of Physics and State Key Laboratory of Surface Physics, Fudan University, Shanghai 200438, China}
\affiliation{Shanghai Branch, Hefei National Laboratory, Shanghai 201315, P.R. China}

\begin{abstract}
    We investigate the ground states of the $S = 1/2$ staircase $J$-$Q_3$ model in the maximally anisotropic limit by employing projector quantum Monte Carlo simulations. To overcome boundary-induced finite-size ambiguities inherent in the study of spatially modulated structures, we implement a $45^{\circ}$ tilted periodic boundary condition that eliminates intermediate phases and provides direct access to winding-sector transitions of the system. By defining a domain wall density to quantify the spatial modulation of the helical valence bond phase, we perform thermodynamic extrapolations and demonstrate that both the domain wall density and the characteristic wavevector evolve continuously with the coupling ratio, exhibiting no commensurate lock-in behavior. Our results establish that the helical valence bond phase is a genuine two-dimensional incommensurate phase with long-range bond-bond order in the thermodynamic limit, clarifying that winding-sector transitions are finite-size effects enforced by boundary commensurability. Furthermore, we determine the phase transition point between columnar valence bond solid phase and helical valence bond phase to be $g_c = 0.046(2)$.
\end{abstract}

\maketitle

\section{Introduction}

The exploration of quantum phase transitions (QPTs) and exotic states of matter constitutes a cornerstone of modern condensed matter physics~\cite{sachdev2011quantum, sondhi1997continuous}. In conventionally ordered phases, symmetry breaking typically manifests as uniform or commensurate long-range order~\cite{anderson1972more, chaikin1995principles, landau2013statistical}. However, specific competing interactions or geometric constraints may stabilize spatially modulated structures, leading to a variety of commensurate and incommensurate phases~\cite{pokrovsky1979incommensurate_2D, fisher1980infinitely, villain1981two, bak1982incommensurate, selke1988annni}. An incommensurate phase, characterized by a modulation wavevector that forms an irrational ratio with the underlying reciprocal lattice vectors, breaks the discrete translational symmetry of the original lattice and introduces unique collective excitations, such as phasons and amplitudons~\cite{overhauser1971observability, lee1974conductivity, mcmillan1976incommensurate, blinc1981magnetic}. Such phenomena are realized in many strongly correlated systems, including frustrated insulators and twisted moir\'e superlattices~\cite{woods2014commensurate, starykh2015unusual}. Understanding the mechanisms that stabilize these incommensurate states, as well as the nature of the commensurate-incommensurate (C-IC) transitions, remains a fundamental challenge due to the interplay of geometric frustration, chiral perturbations and quantum fluctuations~\cite{whitsitt2018quantum, chepiga2020floating, nyckees2022commensurate}.

In quantum magnetism, strongly correlated spin systems on two-dimensional lattices provide an ideal playground for investigating these unconventional phases~\cite{sachdev2011quantum}. Over the past decades, extensive research has demonstrated that the combination of low dimensionality and competing interactions can give rise to a broad spectrum of novel ground states, ranging from valence bond crystals to exotic spin liquids~\cite{savary2017quantum, zhou2017quantum}. While quantum Monte Carlo (QMC) simulations serve as a primary tool to explore such models~\cite{sandvik2010SSE}, identifying the true nature of modulated states remains highly non-trivial. When dealing with spatially modulated structures, finite-size numerical approaches often struggle to distinguish between a smooth analytic behavior of the modulation period and a complete or incomplete devil's staircase of locked commensurate phases~\cite{bak1982incommensurate, selke1988annni, zhang2011monte}. Although certain infinite tensor network algorithms can directly target the thermodynamic limit, they inherently suffer from variational biases~\cite{jordan2008iPEPS}. To circumvent such ansatz-dependent limitations, unbiased quantum Monte Carlo (QMC) methods remain a primary alternative. However, these exact approaches rely on lattice with finite sizes~\cite{sandvik2019SSE}. Subtleties such as the intrinsic length scales and boundary conditions can introduce severe finite-size ambiguities, making the extraction of true thermodynamic properties a formidable task~\cite{zhao2020HVB}.

\begin{figure}[b]
    \centering
    \begin{minipage}[c]{0.15\textwidth}
        \centering
        \includegraphics[width=\textwidth]{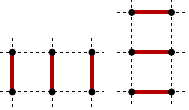}
        \small (a)
    \end{minipage}
    \hspace{2em}
    \begin{minipage}[c]{0.2\textwidth}
        \centering
        \includegraphics[width=\textwidth]{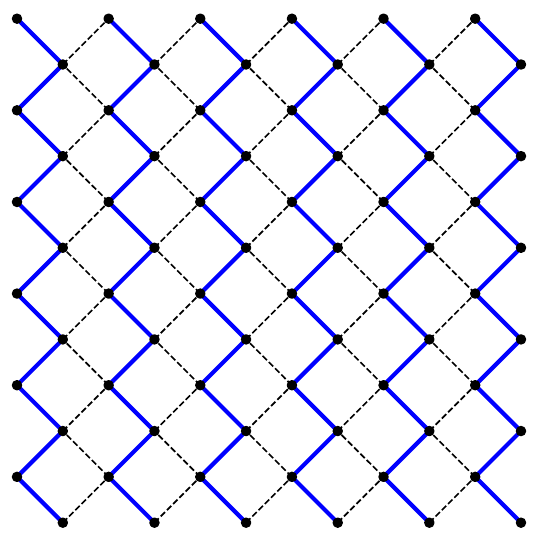}
        \small (b)
    \end{minipage}
    \caption{(a) Schematic illustration of the $Q$ terms. (b) Schematic of the $L \times L \times 2$ lattice geometry with $45^{\circ}$ tilted periodic boundary conditions. Thick blue lines and dashed black lines represent the strong ($J(1 + h)$) and weak ($J(1 - h)$) bonds, respectively.}
    \label{fig:q_term}
    \label{fig:lattice}
\end{figure}

To address these challenges within a concrete numerical setting, we investigate the $J$-$Q_3$ model with the staircase modulation. This model hosts a spatially rotating helical valence bond (HVB) phase. By employing a $45^{\circ}$ tilted periodic boundary condition, we successfully eliminate boundary-induced intermediate phases. Through a systematic analysis of ground-state level crossings and finite-size scaling of the intensive domain wall density, we provide evidence that the HVB phase is a two-dimensional incommensurate phase with long-range order in the thermodynamic limit. Our work demonstrates a robust approach for extracting true thermodynamic properties of spatially modulated quantum states from finite-size analysis.

\section{Methodology}

\subsection{Model}

We study the $S = 1/2$ staircase $J$-$Q_3$ model on a two-dimensional square lattice, originally introduced by Sandvik \textit{et al.}~\cite{zhao2020HVB}. This model is a deformation of the standard $J$-$Q_3$ model~\cite{sandvik2007JQ2, lou2009JQ3_SUN}. The Hamiltonian is given by
\begin{equation}
    H = - \sum_{\langle ij \rangle} J_{ij} \hat{P}_{ij} - Q \sum_{\langle \langle ijklmn \rangle \rangle} \hat{P}_{ij} \hat{P}_{kl} \hat{P}_{mn},
\end{equation}
where $\langle ij \rangle$ denotes the nearest neighbor and $\hat{P}$ is the singlet projector
 \begin{equation}
    \hat{P}_{ij} = \frac{1}{4} - \boldsymbol{S}_i \cdot \boldsymbol{S}_j.
\end{equation}
The $Q$ terms represent products of three parallel singlet projectors that favor columnar valence bond solid (cVBS) order, arranged as illustrated in Fig.~\ref{fig:q_term}(a). The staircase modulation is introduced through the nearest neighbor exchange interactions $J_{ij}$ where the couplings alternate as $J(1+h)$ and $J(1-h)$ to form zigzag staircase chains, as shown in Fig.~\ref{fig:lattice}(b). In this work, we focus exclusively on the $h = 1$ situation, where the exchange interactions are maximally anisotropic: the ``strong" blue bonds have an effective strength of $2J$, while the ``weak" black bonds vanish entirely. We define the coupling ratio as $g = J / (J + Q)$ to tune the system.

\begin{figure}[t]
    \centering
    \begin{minipage}[c]{0.2\textwidth}
        \centering
        \includegraphics[width=\textwidth]{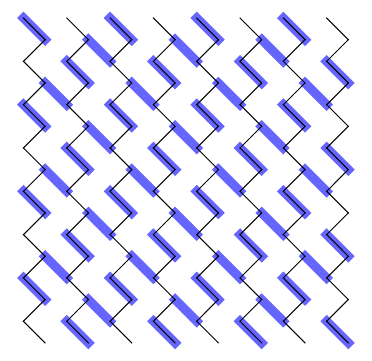}
        \small (a)
    \end{minipage}
    \hspace{10pt}
    \begin{minipage}[c]{0.2\textwidth}
        \centering
        \includegraphics[width=\textwidth]{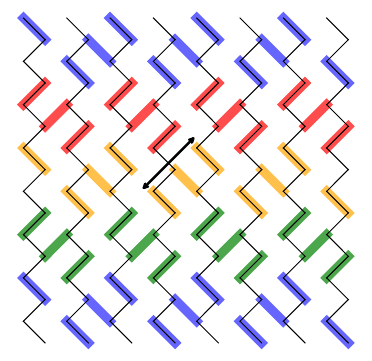}
        \small (b)
    \end{minipage}
    \\
    \begin{minipage}[c]{0.2\textwidth}
        \centering
        \includegraphics[width=\textwidth]{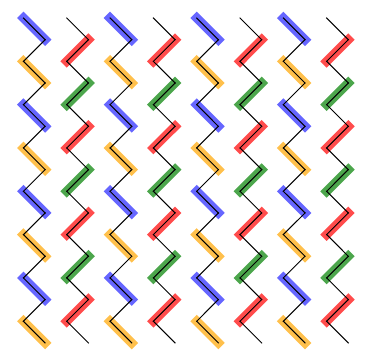}
        \small (c)
    \end{minipage}
    \hspace{12pt}
    \begin{minipage}[c]{0.18\textwidth}
        \centering
        \includegraphics[width=\textwidth]{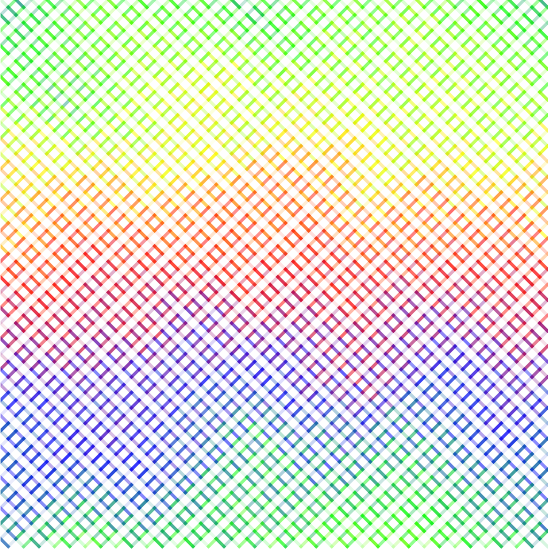}
        \small (d)
    \end{minipage}
    \caption{Schematic illustrations of VBS configurations corresponding to different domain wall densities $\rho$. (a) The completely uniform columnar VBS state with no domain walls ($\rho = 0$). (b) An HVB state with a domain wall density of $\rho = 1/3$, where each uniform domain has a width of exactly three lattice constants along the underlying lattice direction, as highlighted by the double-headed arrow. (c) The maximally dense configuration of HVB phase ($\rho = 1$), where domain walls are packed in the most compact manner and all singlets reside exclusively on the strong bonds. (d) Snapshot of the bond configuration during QMC simulations at $g = 0.17$ and $L = 36$, averaged over 2000 Monte Carlo steps. The opacity of each bond represents the spin correlation, while its color denotes the phase of the local order parameter $T_{\hat{x}} + i T_{\hat{y}}$.}
    \label{fig:columnar}
    \label{fig:helical}
    \label{fig:dense}
\end{figure}

At the isotropic limit $h = 0$, the model reduces to the standard $J$-$Q_3$ model, which exhibits a very weak first-order phase transition between the N\'eel state and the columnar VBS phase~\cite{lou2009JQ3_SUN, sen2010staggered_first, sandvik2012JQ3_first, shu2022JQ3_first, wang2022first, zhao2022first, deng2024first, d2024first, zhu2026first, zhang2026JQ3_first}. This transition point is widely considered to lie in close proximity to a deconfined quantum critical point (DQCP)~\cite{senthil2004DQCP, sandvik2007JQ2, melko2008JQ2_DQCP, chen2013DQCP_first, nahum2015DQCP, nahum2015DQCP2, shao2016JQ2_DQCP, qin2017DQCP, ma2018DQCP, takahashi2024DQCP, song2025DQCP}. For $h > 0$, the HVB phase emerges from this transition point~\cite{zhao2020HVB}. Within the HVB phase, the system continuously rotates among the four degenerate columnar VBS configurations along the direction of the staircase chains (i.e., the $y$-boundary direction). A sketch of HVB configuration is shown in Fig.~\ref{fig:helical}(b).

To characterize the nature of the HVB phase, we employ a specific lattice geometry. The system is defined on a $L \times L \times 2$ lattice to implement $45^{\circ}$ tilted periodic boundary condition, depicted in Fig.~\ref{fig:lattice}(b). This tilted boundary condition is essential for stabilizing HVB phase near the transition points between different winding sectors, which we discuss in detail in subsequent sections.

\subsection{Method}

We use the projector stochastic series expansion (SSE) quantum Monte Carlo (QMC) algorithm~\cite{sandvik1999SSE, sandvik2010SSE, sandvik2019SSE, sandvik2005projector, sandvik2010projector} in the valence bond basis~\cite{beach2006VBbasis}. Rather than applying a large power of Hamiltonian, $(-H)^m$, to a trial state, we use the operator $\exp (-\beta H)$ in the limit $\beta \rightarrow \infty$. This approach allows us to perform calculations in valence bond basis while benefiting from the parallelization efficiency inherent to projector QMC. Furthermore, this choice provides a direct estimate of the ground state energy $E = - \langle n \rangle / \beta$ identical to the one used in standard SSE, where $\langle n \rangle$ is the average expansion power. This method may avoid the computationally demanding task of evaluating six-spin correlation functions for the $Q$ terms, enabling simulations of system sizes up to $L = 96$.

\begin{figure*}[!t]
    \centering
    \includegraphics[width=1.0\textwidth]{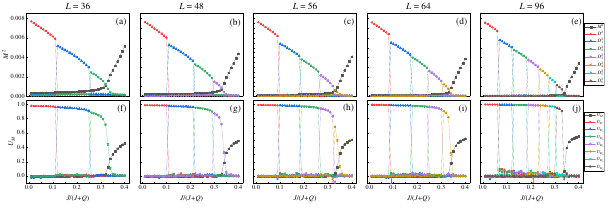}
    \caption{Order parameters and corresponding Binder cumulants as a function of the coupling ratio $g = J/(J+Q)$. Panels (a) to (e) show the order parameters for system sizes $L = 36, 48, 56, 64$, and $96$, respectively, while panels (f) to (j) present their respective Binder cumulants.}
    \label{fig:orderParameter}
\end{figure*}

\subsection{Domain Wall Density}

To quantify the spatial modulation of HVB phase, we define the domain wall density $\rho$ through the correlation functions of the system:
\begin{equation}
    \rho = \frac{C_s - C_w}{C_s + C_w},
\end{equation}
where 
\begin{equation}
    C_s = \sum_{\langle ij \rangle \in \mathrm{strong}} \langle \boldsymbol{S}_i \cdot \boldsymbol{S}_j \rangle
\end{equation}
denotes the sum of correlation functions over all nearest neighbor strong bonds, and
\begin{equation}
    C_w = \sum_{\langle ij \rangle \in \mathrm{weak}} \langle \boldsymbol{S}_i \cdot \boldsymbol{S}_j \rangle
\end{equation}
is the corresponding sum over the weak bonds.

The physical meaning of the domain wall density can be intuitively understood through the spatial arrangement of the fourfold degenerate columnar VBS patterns. As schematically illustrated in Fig.~\ref{fig:columnar}(a), (b) and (c), the underlying lattice is represented by thin black lines with a coupling strength of $2J$, while the vanishing weak bonds are omitted from the sketch. The thick bonds of distinct colors represent the VBS configurations, corresponding to the four degenerate local orientations. A domain wall is physically realized when the singlet arrangement shifts between these different VBS orientations in adjacent regions.

In the columnar state shown in Fig.~\ref{fig:columnar}(a), the system is completely uniform with no domain walls present at all. The singlets are distributed uniformly across the lattice. Exactly half of the singlets lie on the strong bonds and the other half on the weak bonds, thereby yielding $\rho = 0$. Conversely, in the highly modulated HVB phase shown in Fig.~\ref{fig:dense}(c), the domain walls are arranged in the most compact manner possible. All singlets in this configuration are on the strong bonds, maximizing the domain wall density to its upper bound of $\rho = 1$. These two scenarios represent the extreme limits of the system. Fig.~\ref{fig:helical}(b) shows an example lying between these two limits. In this state, each domain of a local orientation has a width of exactly three lattice constants along the underlying lattice direction, as highlighted by the double-headed arrow. In this case, the domain wall density is inversely proportional to this width, evaluating precisely to $\rho = 1/3$. Broadly speaking, $\rho$ corresponds to the inverse of the average width of each uniform pattern, measured along the underlying lattice direction rather than the tilted $y$-boundary direction.

\section{Transitions between Winding Sectors}

To investigate the phase transitions systematically, we introduce the order parameters and Binder cumulants for each phase. For the N\'eel state, the staggered magnetization is defined as
\begin{equation}
    m_z = \frac{1}{N} \sum_{i} (-1)^{x_i + y_i} S_i^z,
\end{equation}
where $N = 2 L^2$ represents the total number of lattice sites, and $x_i$, $y_i$ denote the spatial coordinates of site $i$ defined with respect to the original underlying square lattice axes rather than the tilted direction of the periodic boundary conditions. The associated Binder cumulant is given by
\begin{equation}
    U_{M} = \frac{3}{2} - \frac{\langle m_z^4 \rangle}{2 \langle m_z^2 \rangle^2}.
\end{equation}
To characterize the VBS and HVB phases, we first define the local order parameter centered at site $i$ as
\begin{equation}
    \begin{aligned}
        T_{\hat{x}}(\boldsymbol{r}_i) = \frac{1}{6} (-1)^{x_i} & \big( S_{\boldsymbol{r}_i}^z S_{\boldsymbol{r}_i + \hat{x}}^z + S_{\boldsymbol{r}_i + \hat{y}}^z S_{\boldsymbol{r}_i + \hat{x} + \hat{y}}^z \\
        & + S_{\boldsymbol{r}_i - \hat{y}}^z S_{\boldsymbol{r}_i + \hat{x} - \hat{y}}^z - S_{\boldsymbol{r}_i}^z S_{\boldsymbol{r}_i - \hat{x}}^z \\
        & - S_{\boldsymbol{r}_i + \hat{y}}^z S_{\boldsymbol{r}_i - \hat{x} + \hat{y}}^z - S_{\boldsymbol{r}_i - \hat{y}}^z S_{\boldsymbol{r}_i - \hat{x} - \hat{y}}^z \big),
    \end{aligned}
\end{equation}
and $T_{\hat{y}}(\boldsymbol{r}_i) = T_{\hat{x}}(\boldsymbol{r}_i)(\hat{x} \leftrightarrow \hat{y})$, where the spatial coordinates $\boldsymbol{r}_i = (x_i, y_i)$ and unit vectors $\hat{x}$, $\hat{y}$ are defined with respect to the original underlying square lattice axes~\cite{zhao2020HVB}. We then consider the helical order parameter
\begin{equation}
    D_n = \sum_{i} \big[ T_{\hat{x}}(\boldsymbol{r}_i) + i T_{\hat{y}}(\boldsymbol{r}_i) \big] e^{-i \boldsymbol{k}_n \cdot \boldsymbol{r}_i},
\end{equation}
where $n$ is the winding number, and
\begin{equation}
    \boldsymbol{k}_n = \left( \frac{n\pi}{L}, \frac{n\pi}{L} \right)
\end{equation}
denotes the wavevector parallel to the $y$-boundary direction. The order parameter corresponding to the columnar VBS phase is $D_0$, which we denote simply as $D$. The corresponding Binder cumulants for the VBS and HVB phases are defined as
\begin{equation}
    U_{D_{n}} = 2 - \frac{\langle D_n^4 \rangle}{\langle D_n^2 \rangle^2}.
\end{equation}
In the HVB phase, the system can be partitioned into distinct winding sectors, each identified by $n$. This winding number represents the phase rotation of the complex VBS order parameter. Specifically, it equals the total phase change upon wrapping around the system once along the $y$-boundary direction, divided by $2\pi$. Due to the fourfold degeneracy of the VBS pattern, a full $2\pi$ rotation sequentially traverses all four orientations.

We calculate these order parameters and Binder cumulants, as illustrated in Fig.~\ref{fig:orderParameter}. As the coupling ratio $g$ increases, the system first undergoes a phase transition from the columnar VBS phase to the HVB phase, entering the $n = 1$ winding sector. Within the HVB phase, a series of winding-sector transitions occur, where the winding number increments by one each time. The number of accessible winding sectors increases with system size $L$, with up to six sectors for $L = 96$. The order parameter $D_n$ reveals that the bond-bond order of the system gradually decreases as $g$ increases. Eventually, upon further increasing $g$, the system enters the N\'eel state from the HVB phase. As indicated by the behavior of the Binder cumulants, this HVB-N\'eel transition exhibits a similar characteristics to the columnar VBS-N\'eel transition in the standard $J$-$Q_3$ model.

The properties of these transitions depend on the boundary conditions at finite sizes. In previous studies employing standard periodic boundary conditions, the emergence of intermediate states is observed at the transition points where the wavevector $\boldsymbol{k}$ becomes parallel to the boundary direction~\cite{zhao2020HVB}. In contrast, the $45^{\circ}$ tilted periodic boundary conditions employed here completely eliminate such intermediate phases. This phenomenon can be understood intuitively: under the tilted boundary conditions applied here, the staircase and boundary directions align along the same direction.

\begin{figure}[t]
    \centering
    \includegraphics[width=0.35\textwidth]{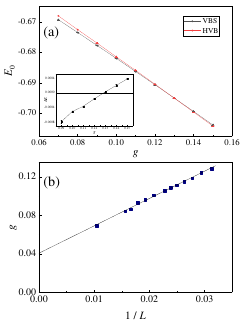}
    \caption{(a) Ground-state energy curves of the cVBS phase and the HVB phase near the transition point for $L = 32$, where the inset plots the energy difference $\Delta E$ between the two phases. (b) Extrapolation of the transition points using the crossing points of energy curves to the thermodynamic limit. The linear fit result yields a transition point at $g_c = 0.04$.}
    \label{fig:transition}
\end{figure}

Due to the intrinsic sampling dynamics of the QMC algorithm, the system can easily become trapped in metastable states near the transition points. This causes the simulation to temporarily remain in the state belonging to an adjacent winding sector even after passing the transition point, as also observed in previous studies~\cite{zhao2020HVB}. By tracking and comparing the ground state energy curves of the two competing sectors near the transition points, as shown for $L = 32$ in Fig.~\ref{fig:transition}(a), we find that the first derivative of the energy curves exhibits a discontinuity at the crossing point. This behavior come from the direct intersection of distinct energy levels belonging to separate winding sectors, forming an actual level crossing. Although this intersection indicates that the transition is first-order, there is no phase coexistence. Since a state cannot mathematically possess two different winding numbers simultaneously, the transition occurs without phase coexistence even in thermodynamic limit.

In fact, this transition consists of two different phenomena: a finite-size winding-sector transition from $n = 0$ to $n = 1$, and the cVBS-HVB phase transition from $\boldsymbol{k} = 0$ to $\boldsymbol{k} \neq 0$. In the thermodynamic limit, the former winding-sector transition no longer exists. This argument will be further discussed in the next section. Within our current finite-size numerical simulations, we do observe a change in the slopes of the two energy curves at the crossing point. However, it remains challenging to determine whether this behavior persists in the thermodynamic limit, as this phenomenon may merely be a finite-size effect associated with the sector transition. Since the cVBS-HVB transition is fundamentally a transition from a commensurate phase to an incommensurate phase, other scenarios for the transition type cannot be entirely ruled out~\cite{mcmillan1976incommensurate, schulz1980incommensurate_2D, bak1982incommensurate}.

Finally, without any interference from intermediate states, the VBS-HVB transition point of various lattice sizes $L$ can be identified by tracking the energy crossing point. By performing a finite size scaling extrapolation to the thermodynamic limit via a linear fit shown in Fig.~\ref{fig:transition}(b), we find the VBS-HVB transition point to be $g_c = 0.04$.

\section{Extrapolation of Domain Wall Density}

\begin{figure}[t]
    \centering
    \includegraphics[width=0.35\textwidth]{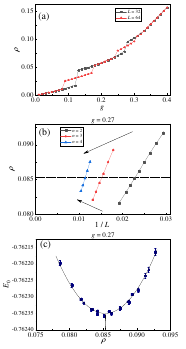}
    \caption{(a) Domain wall density $\rho$ as a function of coupling ratio $g$ for $L = 32$ and $64$. (b) Behavior of $\rho$ versus $1 / L$ at a fixed $g = 0.27$. The horizontal dashed line denotes the thermodynamic limit value ($\rho = 0.08531(5)$) extracted from the fit in (c). (c) The ground-state energy $E_0$ as a function of $\rho$ at $g = 0.27$. The solid curve is a quadratic fit and the vertical dashed line marks the energy minimum.}
    \label{fig:rho}
\end{figure}

To further clarify the nature of the VBS phase, we analyze the behavior of the domain wall density $\rho$ defined in Sec.~II. Fig.~\ref{fig:rho}(a) displays $\rho$ as a function of the coupling ratio $g$ for lattice sizes $L = 32$ and $64$. The curves exhibit step-like features, where each individual plateau corresponds to a sector characterized by the winding number $n$, with $n = 0$ representing the cVBS phase. Interestingly, the $\rho$ curve for the $n = 2$ sector of $L = 64$ overlaps with that of the $n = 1$ sector of $L = 32$. This collapse demonstrates that $\rho$ behaves as an intensive quantity, whereas the winding number $n$ scales extensively with the system size. Consequently, the discrete winding number $n$ is not suitable for describing the physical states of the system in the thermodynamic limit.

Notably, these steps tilt slightly upward as $g$ increases within any given winding sector. This behavior stems from transient fluctuations in the system. The deconfined quantum criticality of the $J$-$Q_3$ model has fractionalized $S = 1/2$ spinon excitations. When a finite staircase modulation $h \neq 0$ breaks the explicit lattice symmetries, an emergent U($1$) symmetry persists within the coarse-grained order parameters of the HVB phase. Transitions between different winding sectors in the HVB phase are microscopically mediated by the creation and subsequent annihilation of these spinon pairs~\cite{tang2011spinon, sandvik2011spinon, tang2013spinon, shao2015domain_wall, suwa2016spinon, liu2018spinon}. As $g$ increases, the system favors the temporary excitation of these spinons. When the two spinons within a pair become separated by a certain distance, several domain walls are formed between them. The contributions from these transiently excited domain walls are captured by $\rho$.

Fixing the coupling ratio and varying the system size allows us to track the finite size behavior of $\rho$, as illustrated in Fig.~\ref{fig:rho}(b). We here fix $g = 0.27$. Data points connected by the same line belong to the same winding sector. Within a given sector, $\rho$ decreases gradually with increasing $L$. Upon reaching a spatial threshold, the system enters the next winding sector, accompanied by a discontinuous increase in $\rho$. By adjusting $L$ at a fixed $g = 0.27$, we also obtain a collection of pairs of $\rho$ and their corresponding ground state energy $E_0$, as plotted in Fig.~\ref{fig:rho}(c). In our QMC simulations, we apply periodic boundary conditions that force the system to be commensurate. In the thermodynamic limit, however, the system should be commensurate with arbitrary wavevector $\boldsymbol{k}$. Thus, $\rho$ in the thermodynamic limit must correspond to the minimum of the ground state energy. We perform a quadratic polynomial fit to $E_0$ yields $\rho = 0.08531(5)$ at $g = 0.27$. As a consistency check, this extrapolated $\rho$ is plotted back onto Fig.~\ref{fig:rho}(b) as a horizontal dashed line.

\begin{figure}[t]
    \centering
    \includegraphics[width=0.35\textwidth]{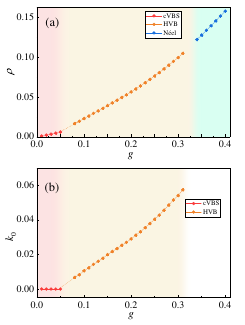}
    \caption{The extrapolated thermodynamic limit of (a) the domain wall density $\rho$ and (b) the characteristic wavevector magnitude $k_0$ as a function of $g$ across the VBS, HVB and N\'eel phases. In (a), the different colored shaded regions distinguish the separate phases, while the yellow shaded region in (b) highlights the regime where the wavevector $\boldsymbol{k}$ remains well-defined. The dashed lines represent linear fits to the HVB data points. The cVBS-HVB phase transition point is $g_c = 0.046(2)$.}
    \label{fig:extrapolation}
\end{figure}

Applying this procedure across all sampled $g$ points in the HVB phase yields the full extrapolation result shown in Fig.~\ref{fig:extrapolation}. Since the wavevector is given by $\boldsymbol{k} = (n\pi / L, n\pi / L)$, we define $k_0 = n / L$. Both the extrapolated domain wall density $\rho$ and the wavevector $k_0$ evolve continuously with $g$ in the HVB phase. We do not observe any evidence of $\rho$ or $k_0$ being locked-in at some specific commensurate values in the extrapolated thermodynamic limit. Instead, available numerical evidence suggests that $k_0$ varies smoothly with $g$, implying this state could be either a ``floating" phase or an incommensurate phase with long-range order. In a typical ``floating" phase, however, the characteristic wavevector is expected to be broadened rather than forming a $\delta$-peak, and the correlation functions should exhibit a power-law decay. Combined with the existence of long-range bond order in the HVB phase, our results support that the HVB phase is an incommensurate phase with long-range order~\cite{mcmillan1976incommensurate, pokrovsky1979incommensurate_2D, bak1982incommensurate, schulz1980incommensurate_2D, beccaria2007floating}, and winding-sector transitions are finite size effects arising from the commensurability enforced by the periodic boundary conditions.

In practice, the HVB data in close proximity to the cVBS-HVB transition cannot be directly obtained through numerical simulations, as they correspond to an extremely small wavevector $\boldsymbol{k}$ that requires system sizes beyond available computational capabilities. Therefore, we employ a linear extrapolation, shown as dashed lines in Fig.~\ref{fig:extrapolation}. The result reveals a cVBS-HVB transition point of $g_c = 0.046(2)$, which is in agreement with the value determined via the energy crossing points in the previous section. Finally, as $g$ increases further and approaches the N\'eel phase boundary, the wavevector $\boldsymbol{k}$ ceases to be well-defined due to fluctuations. Consequently, the extrapolation in the vicinity of the HVB-N\'eel transition also becomes practically unobtainable.

\section{Conclusion}

In conclusion, we have investigated the ground state phase diagram of the staircase $J$-$Q_3$ model in the maximally anisotropic limit ($h = 1$). By implementing $45^{\circ}$ tilted periodic boundary conditions, we have successfully eliminated the intermediate states that appear under standard periodic boundary conditions, thereby enabling a direct observation of winding-sector transitions. To effectively track the spatial modulation of the HVB phase, we introduce the domain wall density $\rho$. Through an extrapolation analysis of the ground-state energy crossing points between competing winding sectors, the phase transition point from the columnar VBS to the HVB phase is $g_c = 0.046(2)$. However, the exact nature of this transition cannot be definitively determined from QMC simulations alone. Since the QMC simulations employed here can only access finite system sizes, the results are inevitably intertwined with finite-size winding-sector transitions. Given that these discrete winding sectors no longer exist in the thermodynamic limit, a full exploration of this transition requires complementary theoretical approaches, which we leave for future investigations.

Crucially, our thermodynamic extrapolations demonstrate that both the domain wall density $\rho$ and the wavevector $\boldsymbol{k}$ evolve continuously with $g$ as the system size approaches infinity.  This result establishes that the HVB phase is a two-dimensional incommensurate phase with bond-bond long-range order in the thermodynamic limit. All sectors with $n \geq 1$ and the transitions between them arise from boundary-enforced commensurability.

\begin{acknowledgments}
This work is supported by the National Key Research and Development Program of China Grant No. 2022YFA1404204, and the National Natural Science Foundation of China Grant Nos. 12274086 and 12564021, and the Quantum Science and Technology-National Science and Technology Major Project (Grant No. 2024ZD0300104).

\end{acknowledgments}

\bibliography{HVB}

\end{document}